\documentclass[conference]{IEEEtran}
\IEEEoverridecommandlockouts
\usepackage{cite}
\usepackage{amsmath,amssymb,amsfonts}
\usepackage{algorithmic}
\usepackage{graphicx}
\usepackage{textcomp}
\usepackage{xcolor}
\usepackage{url}
\def\BibTeX{{\rm B\kern-.05em{\sc i\kern-.025em b}\kern-.08em
    T\kern-.1667em\lower.7ex\hbox{E}\kern-.125emX}}
\begin{document}

\title{Towards LLM-Powered Task-Aware Retrieval of Scientific Workflows for Galaxy\\
}

\author{Shamse Tasnim Cynthia \hspace{4mm}  Banani Roy\\
\normalsize Department of Computer Science, University of Saskatchewan, Canada\\
\normalsize \{shamse.cynthia, banani.roy\}@usask.ca
}
\maketitle

\begin{abstract}
Scientific Workflow Management Systems (SWfMSs) such as Galaxy have become essential infrastructure in bioinformatics, supporting the design, execution, and sharing of complex multi-step analyses. Despite hosting hundreds of reusable workflows across domains, Galaxy’s current keyword-based retrieval system offers limited support for semantic query interpretation that often fails to surface relevant workflows when exact term matches are absent. To address this gap, we propose a task-aware, two-stage retrieval framework that integrates dense vector search with large language model (LLM)–based reranking. Our system first retrieves candidate workflows using state-of-the-art embedding models, then reranks them using instruction-tuned generative LLMs (GPT-4o, Mistral-7B) based on semantic task alignment. To support robust evaluation, we construct a benchmark dataset of Galaxy workflows annotated with semantic topics via BERTopic and synthesize realistic task-oriented queries using LLMs. We conduct a comprehensive comparison of lexical, dense, and reranking models using standard IR metrics, presenting the first systematic evaluation of retrieval performance in the Galaxy ecosystem. Results show that our approach significantly improves top-k accuracy and relevance, particularly for long or under-specified queries. We further integrate our system as a prototype tool within Galaxy, providing a proof-of-concept for LLM-enhanced workflow search. This work advances the usability and accessibility of scientific workflows, especially for novice users and interdisciplinary researchers.

\end{abstract}

\begin{IEEEkeywords}
scientific workflow management systems, large language models, task-aware, natural language query
\end{IEEEkeywords}

\section{Introduction}
Scientific Workflow Management Systems (SWfMSs) such as Galaxy~\cite{afgan2018galaxy} have become indispensable tools in bioinformatics, enabling researchers to design, execute, and share multi-step analyses through visually composed pipelines. The Galaxy platform, in particular, hosts hundreds of reusable workflows across diverse domains, including RNA-seq, ChIP-seq, metagenomics, and proteomics\footnote{\url{https://usegalaxy.org/workflows/list_published}}. However, as the number of publicly available workflows continues to grow, effectively identifying an appropriate workflow based on a user's high-level biological task or intent remains a major challenge \cite{gu2023sworts}.

Currently, Galaxy's retrieval capabilities rely heavily on keyword matching over workflow titles and metadata. These lexical methods suffer from well-known limitations such as vocabulary mismatch and poor support for semantic reasoning, often resulting in suboptimal user experience~\cite{leipzig2017review, gu2024proswats}. For example, a user query like "align single-end reads using HISAT2" may fail to retrieve relevant workflows if exact term matches are not present in the metadata. This gap limits the discoverability and reuse of workflows, especially for novice users or interdisciplinary researchers unfamiliar with precise tool names or formats.
To address these limitations, this study investigates a broad range of retrieval models—including lexical, dense vector, and LLM-based reranking approaches to identify the most effective method for task-aware workflow retrieval in Galaxy. 

Recent advances in neural information retrieval (IR) have introduced dense embedding models that encode queries and documents into semantically meaningful vector spaces~\cite{karpukhin2020dense, reimers2019sentence}, making them particularly suitable for the scientific domain where queries are often long, descriptive, and diverse in terminology. This study evaluates these retrieval strategies across general-purpose workflows and curated training workflows, with the goal of improving both the accessibility and relevance of retrieved workflows for Galaxy users.

Prior studies have explored various strategies to enhance system usability and improve workflow discoverability through more effective search mechanisms \cite{gu2024proswats,gu2023plan,liu2010searching,diao2022bioinformatic,starlinger2014similarity}. For example, Gu et al.~\cite{gu2024proswats} proposed a proxy-based retrieval approach that selects representative workflows for each query to bridge the gap between textual expressions and underlying semantic intent. Other works, such as those by Liu et al.~\cite{liu2010searching} and Gu et al.~\cite{gu2023plan}, focused on matching simple keyword-based queries against workflow attributes like titles and module names. Starlinger et al.~\cite{starlinger2014similarity} applied TF-IDF based similarity scoring over workflow descriptions to support retrieval. Additionally, Diao et al.~\cite{diao2022bioinformatic} incorporated structural workflow elements, such as service relationships, to enrich retrieval accuracy. However, to the best of our knowledge no prior studies have leveraged large language models (LLMs) to support task-aware natural language retrieval in Galaxy or similar scientific workflow environments.

\begin{figure*}
    \centering
    \includegraphics[width=0.8\textwidth]{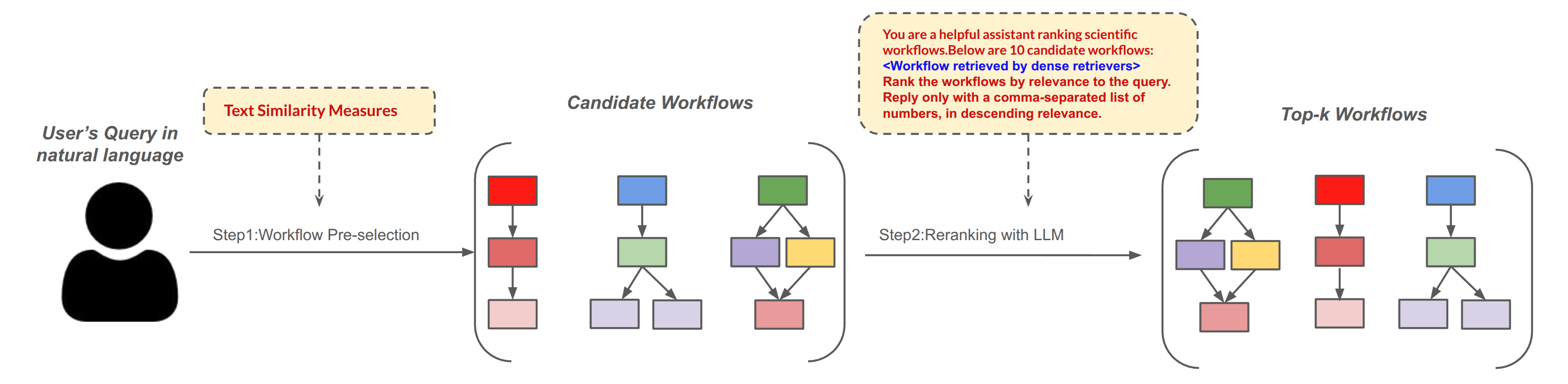}
    \caption{Overview of our proposed framework}
    \label{fig:framework}
\end{figure*}

To address the limitations of keyword-based search in Galaxy, we propose a workflow retrieval approach that combines dense retrieval with large language model (LLM)–based reranking. The primary contributions of this papers are as follows:
\begin{itemize}
    \item We develop a semantic retrieval pipeline that accepts natural language queries and retrieves candidate workflows using state-of-the-art embedding models. These candidates are then reranked using instruction-tuned generative LLMs, enabling fine-grained relevance scoring aligned with user intent.
    \item We construct a high-quality benchmark dataset of Galaxy workflows annotated with semantic topics via BERTopic~\cite{grootendorst2022bertopic}, and generate realistic, task-driven queries using LLMs conditioned on workflow metadata. To the best of our knowledge, this is the first effort to synthesize evaluation-grade natural text queries for scientific workflow retrieval at scale.
    \item We conduct a comprehensive evaluation of lexical, dense, and reranking-based retrieval models using standard information retrieval metrics, presenting the first systematic comparison of workflow retrieval effectiveness within the Galaxy ecosystem. Our results demonstrate that the proposed two-stage approach substantially improves retrieval accuracy and semantic relevance. To further validate its practical utility, we integrate our method as a prototype tool, \textit{WorkflowExplorer} within the Galaxy platform, offering a proof-of-concept for task-aware, LLM-enhanced workflow discovery.
\end{itemize}  

\textbf{Replication Package} is available in our online appendix \cite{replication_package}.



\begin{figure*}
    \centering
    \includegraphics[width=0.8\textwidth]{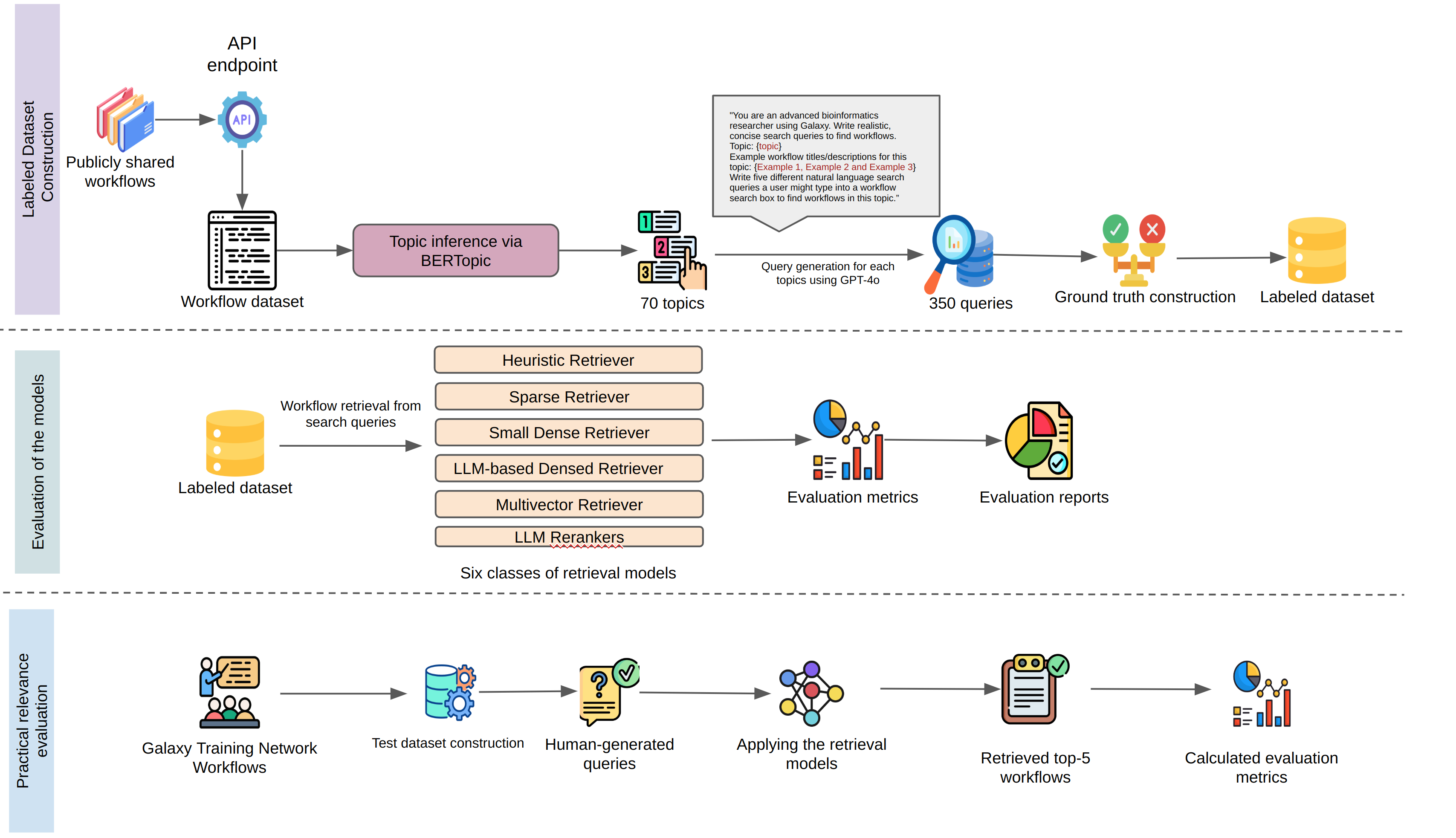}
    \caption{Schematic diagram of the methodology}
    \label{fig:methodology}
\end{figure*}
\section{Related Works}
Prior research in scientific workflow retrieval has explored a range of techniques, from keyword-based matching to structure-aware semantic similarity. For example, Shao et al.~\cite{shao2009wise} developed a keyword-based search engine for workflows collected from Kepler, Taverna, and myGrid. Their system demonstrated strong precision in returning relevant workflows that matched users’ keyword queries, offering a practical baseline for early-stage retrieval tasks.
Starlinger et al. \cite{starlinger2014similarity} at first used bag of words and TF-IDF models to compute the similarity between the query and workflow description. Later in \cite{starlinger2016effective}, they proposed a more structurally sensitive approach that incorporates the execution order of workflow components. By applying topological sorting and normalization techniques, their method enhanced retrieval accuracy for workflows with well-defined procedural sequences. However, this approach requires users to input a complete or partial workflow structure to initiate a query, which limits its accessibility, specially for novice users who lack prior workflow knowledge. Therefore, semantic technologies have been increasingly leveraged to enhance workflow retrieval by capturing deeper relationships beyond keyword matching. For instance, Zhou et al.~\cite{zhou2016scientific} introduced a semantic similarity algorithm that incorporates both hierarchical structure and workflow descriptions to enable more effective clustering and recommendation of scientific workflows. Similarly, Yu et al.~\cite{yu2020workflow} proposed a graph-based embedding approach that represents entire workflows across both transactional and scientific domains by combining structural representations with textual metadata. While these approaches move beyond surface-level lexical matching, they often overlook the full structural richness and task-specific semantics embedded in scientific workflows, which can play a critical role in aligning user intent with workflow functionality.

In a similar direction, Bergmann et al.~\cite{bergmann2014similarity} applied graph representations and an A* search algorithm to compute similarity between semantic workflows. While effective in capturing graph-based structure, their technique was primarily evaluated on small sets of curated workflows and similarly assumes structured input from the user. 

More recently, Wen et al.\cite{wen2020heterogeneous} modeled scientific workflows as heterogeneous information networks (HINs), capturing both data objects and logical relationships, and employed metapath-based similarity techniques for retrieval. Complementary to this, other studies have represented workflows as semantically labeled directed graphs to enhance semantic matching\cite{malburg2018query, bergmann2018similarity}. Although these approaches improve the expressiveness of workflow representation, they often require users to specify detailed substructures or constraints, posing a usability barrier for general or exploratory search scenarios.

While structure-aware and graph-based retrieval methods have significantly advanced the field, a persistent limitation is that these systems often require structured queries, presuppose user familiarity with specific workflow components, or emphasize low-level tool similarity over task intent. Such assumptions create a usability barrier, particularly for users formulating queries in natural language or seeking workflows based on high-level analytical goals. Notably, the potential of LLMs to bridge this semantic gap remains underexplored. Our work directly addresses this shortcoming by proposing a task-aware retrieval framework that leverages LLMs to interpret user intent and retrieve relevant workflows from real-world repositories such as Galaxy.

\section{Design of WorkflowExplorer}
\subsection{Overall Framework}

The design of \textit{WorkflowExplorer}, illustrated in Figure~\ref{fig:framework}, follows a two-stage retrieval architecture inspired by modern neural information retrieval pipelines. It is specifically tailored for the Galaxy ecosystem, aiming to retrieve semantically relevant workflows in response to user-provided natural language queries. The framework consists of two primary stages: (1) Workflow Candidate Retrieval and (2) LLM-based Reranking.

In the first stage, given a user query $Q$, we retrieve a set of top-$k$ candidate workflows $W_c$ using dense vector similarity. These candidates are selected based on the cosine similarity between the query embedding and the embedding representations of workflow metadata (titles and descriptions), computed using state-of-the-art sentence-level encoders (e.g., BGE, SBERT, Ada-002).

In the second stage, the retrieved candidates are reranked using instruction-tuned generative LLMs, specifically GPT-4o and Mistral-7B. These models evaluate the semantic alignment between the original query and each candidate workflow, assigning a relevance score and reordering the list accordingly.

To ensure robustness, we construct a benchmark corpus of Galaxy workflows annotated with semantic topic labels using BERTopic~\cite{grootendorst2022bertopic}, and generate realistic, diverse task queries using LLM-based prompting. Our system supports queries involving multiple steps, tool intent, or biological context, addressing limitations of keyword-based retrieval in Galaxy.

\subsection{Workflow Candidate Retrieval}

In the first phase, the goal is to retrieve a semantically plausible subset of workflows that align with the user’s intent, expressed through a natural language query $Q$. We evaluate both lexical and neural retrieval strategies to compute the similarity between the query and each workflow in the corpus.

For lexical baselines such as BM25, TF-IDF, and Fuzzy Matching, the query $Q$ is tokenized and compared against preprocessed workflow texts (titles, descriptions, and optionally tool names) using traditional term-frequency and fuzzy string matching algorithms. These models rank workflows based on lexical overlap or token-level scoring functions.

For dense and LLM-based retrieval methods—including BGE-Base, DPR, SBERT-miniLM, ada-002, voyage-l2-instruct, LLM2Vec, GritLM-7B, ColBERTv2, and multi-ada-002—we encode the query $Q$ into a fixed-length vector representation $v_Q$ using a pretrained transformer-based encoder. Likewise, each workflow $W_i$ in the corpus is encoded into $v_{W_i}$ using the same or a compatible model.

We use cosine similarity between these embeddings to identify the top-$k$ most relevant workflows:
\[
W_c = \text{TopK}\left( \cos(v_Q, v_{W_i}) \right), \quad \forall W_i \in \text{Corpus}
\]
For dense retrievers, this approximate nearest neighbor search is performed efficiently using FAISS, enabling scalable retrieval over thousands of workflows. In all cases, we retain the top-$k$ candidates (typically $k = 50$) to be passed into the second-stage reranking module.

\subsection{LLM-based Workflow Reranking}

Although dense retrieval provides semantically rich candidates, it does not fully capture query-specific reasoning or multi-step intent. To address this, we apply a reranking phase using GPT-4o and Mistral-7B, both of which are capable of reasoning over longer contexts.

Each workflow candidate’s description is presented to the LLM alongside the original query in a prompt. The LLM is instructed to output a relevance ranking or preference score over the candidates. These scores are then used to reorder the candidate list, yielding a final ranked set of workflows $W_r$:
\[
W_r = \text{Rerank}_{\text{LLM}}(Q, W_c)
\]

The reranker incorporates nuanced biological context, tool usage, and procedural semantics that dense embeddings may overlook. This stage is particularly effective for queries that involve implicit reasoning, tool combinations, or domain-specific jargon.

\subsection{Implementation and Efficiency Considerations}

To balance effectiveness and efficiency, we perform embedding inference and retrieval on GPU using batched inference. The reranking phase supports model fallback (e.g., Mistral-7B for local use, GPT-4o for high-accuracy API calls). All components are modular and integrated into a Galaxy-compatible prototype, enabling real-time workflow search within the platform.

\section{Experiments}

This section presents the methodology of task-aware natural language workflow retrieval in the Galaxy ecosystem. Fig.~\ref{fig:methodology} shows the schematic diagram of the proposed methodology.

\subsection{Workflow Dataset Extraction}

We collected publicly available Galaxy workflows using the Galaxy API endpoint\footnote{\url{https://usegalaxy.org/api/workflows?show_published=True}}. For each workflow, we retrieved its metadata including \texttt{workflow ID, title, description} and the list of the \texttt{tool} names used the analysis steps. We also downloaded the corresponding \texttt{.ga} file, which encodes the full workflow definition. Next, we stored this structured metadata into a JSON file.

\subsection{Topic Inference via BERTopic}

To semantically group the extracted Galaxy workflows by topic, we employed BERTopic \cite{grootendorst2022bertopic}, a topic modeling technique that combines BERT-based embeddings with clustering. We used the \texttt{MiniLM-L6-v2} model from SentenceTransformers\footnote{\url{https://huggingface.co/sentence-transformers/all-MiniLM-L6-v2}} to obtain dense vector representations. BERTopic then clustered these embeddings into latent topics, each representing a coherent scientific task or domain theme. We found 70 topics in total. This process yielded a mapping from topic labels to workflow IDs, along with descriptive statistics summarizing the topic distribution. Each workflow was subsequently annotated with one topic. The enriched metadata was stored in a new JSON file, forming the basis for query generation and ground truth construction.

\subsection{Query Generation Using LLMs}
Wang et al. \cite{wang2023generating} addressed the limitations of using final review titles for screening prioritisation by exploring alternative queries generated at screening time using instruction-based generative language models such as ChatGPT and Alpaca. Following their methodology, we leveraged GPT-4o to generate five natural language queries per discovered topic to simlulate realistic, topic-aware user search queries. Each prompt to the LLM included a topic name and a few example workflow titles and descriptions. The LLM was instructed to output concise search queries that a user might type to retrieve workflows relevant to that topic. Each generated query was saved along with the seed workflow IDs used for prompting. 

\subsection{Ground Truth Construction}
Next, for the supervised evaluation of the models, we constructed a ground truth mapping from each query to its set of relevant workflows using a lightweight semantic filtering approach. We began with the seed workflows that were used to prompt the LLM when generating the query. To expand this set, we included additional workflows from the same topic that exhibited high textual or tool-based similarity to the query, measured using TF-IDF and keyword overlap. Only those workflows passing a similarity threshold were considered relevant. The resulting set was stored under a \texttt{gold\_workflow\_ids} field for each query, forming a labeled dataset for evaluation.

\subsection{Baseline Retrieval Models and Evaluation Metrics}
We followed the STARK benchmark paper \cite{wu2024stark} to evaluate LLMs within a rigorous, domain-diverse retrieval framework aligned with our goal of scientific workflow retrieval. Based on their setup, we evaluated \textit{six} classes of retrieval models as described below:

\begin{itemize}
    \item \textbf{String Similarity-based Heuristic Retriever}: \texttt{Fuzzy matching}~\cite{chaudhuri2003robust} is a heuristic approach for computing string similarity based on token overlap and edit distance. It employs metrics such as the token set ratio to quantify textual similarity between queries and documents, allowing retrieval in cases of paraphrasing, word reordering, or minor typographical errors.

    \item \textbf{Sparse Retriever}: \texttt{TF-IDF} (Term Frequency–Inverse Document Frequency)~\cite{aizawa2003information} is a classical lexical retrieval model that ranks documents based on the relative importance of query terms. It assigns higher weights to terms that occur frequently within a document but are infrequent across the overall corpus. While effective for capturing keyword salience, TF-IDF does not incorporate document length normalization. \texttt{BM25}~\cite{robertson2009probabilistic} extends TF-IDF by incorporating a probabilistic relevance framework. It adjusts term weights based on both inverse document frequency and document length, yielding more robust ranking in diverse text collections.

    \item \textbf{Small Dense Retrievers:} \texttt{BGE-base} \cite{bge}, \texttt{DPR}~\cite{karpukhin2020dense}, and \texttt{SBERT-miniLM} \cite{SBERT} are dense retrieval models that encode both queries and documents into fixed-length vector embeddings. Relevance is determined by computing similarity between embeddings, typically via cosine similarity or inner product. These models provide efficient and scalable baselines for semantic retrieval, offering a strong balance between performance and computational cost. They are frequently used as benchmarks to evaluate the effectiveness of more advanced large language model (LLM)-based retrievers.

    \item \textbf{LLM-based Dense Retrievers:} \texttt{text-embedding-ada-002 (ada-002)} \cite{openai}, \texttt{voyage-large-2-instruct (voyage-l2-instruct)} \cite{openai}, \texttt{LLM2Vec-Meta-Llama-3-8B-Instruct-mntp (LLM2Vec)}~\cite{behnamghader2024llm2vec}, and \texttt{GritLM-7b}~\cite{muennighoff2024generative} are large language model (LLM)-based embedding models that generate dense, semantically rich vector representations. These models are designed to capture contextual nuances in natural language, enabling more effective semantic matching between queries and documents.

    \item \textbf{Multivector Retrievers:} \texttt{multi-ada-002} \cite{openai} and \texttt{ColBERTv2}~\cite{santhanam2021colbertv2} extend traditional embedding approaches by introducing multi-vector representations for improved semantic coverage. Unlike \texttt{ada-002}, which encodes an entire document into a single embedding vector, \texttt{multi-ada-002} segments each document into overlapping chunks and encodes them individually using the same encoder as the query, allowing for better handling of long or multi-faceted inputs. \texttt{ColBERTv2} represents documents at the token level, producing multiple fine-grained embeddings per document to support late interaction and enable more precise semantic matching between query and document terms.
 
    \item \textbf{LLM Rerankers:} \texttt{Mistral-7B} (Mistral-7B-Instruct-v0.2) \cite{mistral} and \texttt{GPT-4o} (gpt-4o-2024-08-06) \cite{gpt} are used as reranking models to improve the precision of initial results retrieved by \texttt{multi-ada-002}. Specifically, we employ \texttt{GPT-4o} and \texttt{Mistral-7B-Instruct} as second-stage rerankers. Given a query and a set of top-$k$ candidate workflows, the LLMs evaluate the semantic alignment between the query and each candidate’s textual metadata. Each workflow is assigned a relevance score (typically normalized between 0 and 1), enabling a more informed reordering of candidates based on task-level intent and contextual fit.

\end{itemize}

\subsection{Evaluation Protocol}

All retrieval systems were evaluated against the labeled dataset using four standard ranking metrics:
\begin{itemize}
    \item \textbf{Hit@\textit{k}}: \textit{Hit@k} measures whether at least one relevant item appears within the top-$k$ results returned by the model. In our evaluation, we report \textit{Hit@1} and \textit{Hit@5}, where \textit{Hit@1} assesses the accuracy of the top-ranked recommendation, and \textit{Hit@5} evaluates the model’s ability to include a relevant result within a broader set of top-5 candidates.

    \item \textbf{Recall@{\textit{k}}} measures the proportion of relevant items retrieved within the top-$k$ results. It reflects the model’s ability to retrieve all relevant candidates, which is especially important in high-recall scenarios where omissions may be costly. In our setting, we use \textit{Recall@50} for synthesized queries, as the number of relevant workflows per query is typically small (often fewer than 20), and top-50 retrieval offers a reasonable upper bound for evaluation.

    \item \textbf{Mean Reciprocal Rank (MRR)} is a metric used to evaluate the average ranking quality of a retrieval model. It is computed as the reciprocal of the rank at which the first relevant item appears in the result list, averaged over all queries. MRR places emphasis on retrieving the first correct answer as early as possible, making it particularly useful in applications where the top-ranked result has the greatest practical impact.

\end{itemize}

Each model’s metrics were averaged across all queries, and results were exported to CSV files for analysis and reproducibility.

\subsection{Generating Test Queries} To support our model's practical relevance, we constructed a strutured dataset of Galaxy workflows from the Galaxy Training Network. For training purposes, the Galaxy developers have provided 463 workflows under 27 topics. We manually crawled the GitHub repository, identifying individual tutorial folders such as \textit{amr-gene-detection}. Since these workflows were provided the Galaxy developers, we considered them as our ground truth answers. For each tutorial, we collected the \textit{.ga} files and the files were parsed using a JSON reader to extract the workflow's name and the set of tools used. Then we extracted the workflow description from the \textit{data-library.yml} file. Finally, each workflow was tagged with the parent topic name to preserve its domain context. 

For query construction, we selected 25 tutorial topics (non-administrative topics) and extracted representative task names or objectives from their metadata and filenames. Next, we prompted GPT-4o to generate three natural language queries per topic that realistically reflect how users might describe the task in everyday language. This process resulted in a total of 75 high-quality, diverse queries. A domain expert with more than 4 years of experience in scientific workflow management and Galaxy-based analysis manually reviewed the generated queries, filtered out ambiguous or low-relevance examples, and ensured that each query was aligned with its intended topic and set of relevant workflows.

\section{Overall Performance}

\subsection{Effectiveness of LLM-Based Embeddings}

Table~\ref{tab:retrieval-results} presents the retrieval performance across all models on the full set of synthesized queries. 

Among the lexical baselines, \textbf{TF-IDF} emerges as the strongest performer, achieving a Hit@1 of 31.43\% and an MRR of 38.52\%. This indicates its robust ability to match keywords and contextually relevant terms in workflow descriptions. \textbf{BM25} and \textbf{Fuzzy Matching} follow closely, with MRR scores of 33.62\% and 30.91\%, respectively. While BM25 benefits from probabilistic term weighting, Fuzzy Matching based on token overlap heuristics, struggles to capture semantic nuance. Overall, lexical models provide reasonably strong baselines and perform well when queries contain domain-specific tool names or exact terminology. However, their inability to generalize beyond surface-level text limits their top-rank accuracy in more abstract queries.

Small Dense Retrievers consistently outperform lexical models across nearly all metrics. Notably, \textbf{SBERT-miniLM} achieves a Hit@1 of 35.00\% and an MRR of 41.34\%, outperforming other single-vector retrievers like \textbf{BGE-base} (MRR = 38.64\%) and \textbf{DPR} (MRR = 21.69\%). Among the LLM-based Dense Retrievers, \textbf{voyage-l2-instruct} stands out as the best single-stage retriever, achieving the highest Hit@1 (37.86\%) and MRR (42.44\%). This demonstrates the value of aligning embedding space with task-level intent and natural language understanding. Compared to \textbf{LLM2Vec} (MRR = 18.35\%) and \textbf{GritLM-7b} (MRR = 25.31\%), Voyage exhibits superior performance, likely due to broader instruction tuning and more effective semantic modeling. LLM2Vec, despite its domain grounding, struggles with recall and top-rank precision, suggesting a misalignment between query and document embeddings. GritLM shows promise but lags behind due to possible training limitations or model scale. LLM-based Dense Retriever models thus provide a meaningful improvement in understanding user intent, especially in open-ended, task-driven queries common in bioinformatics workflow search.

Among the Multivector Retrievers, the \textbf{multi-ada-002} variant achieves the highest Hit@5 (47.14\%) and strong recall (36.99\%), suggesting that using multiple embedding vectors enables better coverage of diverse semantic subspaces. In contrast, \textbf{ada-002} underperforms significantly (MRR = 3.94\%), highlighting the limitations of generic embeddings not fine-tuned for domain-specific search. \textbf{ColBERTv2}, though designed for fine-grained interaction, delivers only moderate gains (MRR = 26.25\%), likely due to implementation approximations (mean pooling rather than late interaction). These results confirm that semantic embeddings provide more flexibility than lexical models but still vary widely in effectiveness based on model architecture and domain alignment.

\begin{table}[ht]
\centering
\caption{Retrieval performance of lexical, dense, and LLM-based models. Metrics are averaged over all queries.}
\label{tab:retrieval-results}
\resizebox{\linewidth}{!}{
    \begin{tabular}{lcccc}
        \hline
        \textbf{Method} & \textbf{Hit@1 (\%)} & \textbf{Hit@5 (\%)} & \textbf{Recall@50 (\%)} & \textbf{MRR (\%)} \\
        \hline
        BM25                 & 27.14 & 42.14 & 25.48 & 33.62 \\
        Fuzzy                & 25.00 & 37.86 & 24.70 & 30.91 \\
        TFIDF                & 31.43 & 46.43 & 32.05 & 38.52 \\
        BGE-base             & 32.14 & 43.57 & 33.10 & 38.64 \\
        DPR                  & 17.14 & 24.29 & 21.65 & 21.69 \\
        SBERT-miniLM         & 35.00 & 46.43 & 35.59 & 41.34 \\
        ada-002              &  0.71 &  5.00 &  2.28 &  3.94 \\
        voyage-l2-instruct   & \textbf{37.86} & 45.71 & \textbf{35.72} & \textbf{42.44} \\
        LLM2Vec              & 15.71 & 20.00 & 12.64 & 18.35 \\
        GritLM-7b            & 19.29 & 30.71 & 19.76 & 25.31 \\
        ColBERTv2            & 20.00 & 33.57 & 25.39 & 26.25 \\
        multi-ada-002        & 35.00 & \textbf{47.14} & 36.99 & 41.32 \\
    \hline
    \end{tabular}
    }
\end{table}

\begin{table}[ht]
\centering
\caption{Retrieval Performance Across Models of Testing Queries}
\begin{tabular}{lcccc}
\hline
\textbf{Method} & \textbf{Hit@1} & \textbf{Hit@5} & \textbf{Recall@50} & \textbf{MRR} \\
\hline
BM25               & 38.7 & 58.7 & 43.7 & 47.7 \\
Fuzzy              & 17.3 & 41.3 & 38.8 & 28.5 \\
TFIDF              & 41.3 & 64.0 & 47.0 & 53.3 \\
BGE-base           & 62.7 & 76.0 & 55.5 & 68.4 \\
DPR                & 18.7 & 60.0 & 45.4 & 37.0 \\
SBERT-miniLM       & 73.3 & 82.7 & 58.6 & 76.9 \\
ada-002            & 9.3 & 22.7 & 33.4 & 20.6 \\
voyage-l2-instruct & 61.3 & 80.0 & 60.0 & 69.3 \\
LLM2Vec            & 24.0 & 40.0 & 42.9 & 33.9 \\
GritLM-7b          & 17.3 & 37.3 & 41.9 & 29.2 \\
multi-ada-002      & 62.7 & 76.0 & 62.2 & 69.1 \\
ColBERTv2          & 33.3 & 53.3 & 43.9 & 42.0 \\
Mistral-7B         & 58.7 & 72.0 & 93.2 & 64.6 \\
GPT-4o             & 61.3 & 76.0 & 80.1 & 67.9 \\
\hline
\end{tabular}
\label{tab:retrieval-results}
\end{table}

\begin{table*}[ht]
\centering
\caption{Top-5 retrieved workflows for the query \textit{"How can I incorporate an annotated bacterial genome for RNA sequencing analysis using galaxy?"}, before and after Mistral-7b reranking.}
\label{tab:qual-case-study}
\resizebox{0.7\textwidth}{!}{
    \begin{tabular}{p{0.45\linewidth} | p{0.45\linewidth}}
    \textbf{Before Reranking (voyage-l2-instruct)} & \textbf{After Reranking (Mistral-7b)} \\
    \hline
    \textit{Run tools to evaluate sequencing data on quality and quantity Evaluate the output of quality control tools Improve the quality of sequencing data Run a series of tool to identify species in bacterial isolate sequencing data Visualize the species abundance} & \textit{Run a series of tool to annotate a draft bacterial genome for different types of genomic components Evaluate the annotation Process the outputs to formate them for visualization needs Visualize a draft bacterial genome and its annotations} \\
    \hline
    \textit{Run a series of tool to annotate a draft bacterial genome for different types of genomic components Evaluate the annotation Process the outputs to formate them for visualization needs Visualize a draft bacterial genome and its annotations} & \textit{Load genome into Galaxy Annotate genome with Prokka View annotations in JBrowse} \\
    \hline
    \textit{Analysis of RNA sequencing data using a reference genome Reconstruction of transcripts without reference transcriptome (de novo) Analysis of differentially expressed genes} & \textit{Analyze of 16S rRNA sequencing data using the mothur toolsuite in Galaxy Using a mock community to assess the error rate of your sequencing experiment Visualize sample diversity using Krona and Phinch} \\
    \hline
    \textit{Check a sequence quality report generated by Falco/MultiQC for RNA-Seq data Explain the principle and specificity of mapping of RNA-Seq data to an eukaryotic reference genome Select and run a state of the art mapping tool for RNA-Seq data Evaluate the quality of mapping results Describe the process to estimate the library strandness Estimate the number of reads per genes Explain the count normalization to perform before sample comparison Construct and run a differential gene expression analysis Analyze the DESeq2 output to identify, annotate and visualize differentially expressed genes Perform a gene ontology enrichment analysis Perform and visualize an enrichment analysis for KEGG pathways} & \textit{Load data (genome assembly, annotation and mapped RNASeq) into Galaxy Perform a transcriptome assembly with StringTie Annotate lncRNAs with FEELnc Classify lncRNAs according to their location Update genome annotation with lncRNAs} \\
    \hline
    \textit{Analysis of RAD sequencing data without a reference genome SNP calling from RAD sequencing data Calculate population genomics statistics from RAD sequencing data} & \textit{Workflow 2: Data Cleaning And Chimera Removal [Galaxy Training: 16S Microbial Analysis With Mothur} \\
    \hline
    \end{tabular}
}
\end{table*}

\subsection{Comaparison with curated training workflows} 
The results reveal that almost all models achieve higher accuracy and semantic alignment on the curated training workflows. For example, \textbf{TF-IDF} improves from 31.43\% to 41.3\% in Hit@1 and from 38.52\% to 53.3\% in MRR, demonstrating that even traditional lexical retrieval benefits from cleaner, well-structured metadata. Similarly, \textbf{BM25} shows notable gains in both Hit@1 (from 27.14\% to 38.7\%) and MRR (from 33.62\% to 47.7\%), indicating that workflow clarity and annotation consistency play an important role in improving sparse retrieval effectiveness.

Dense retrievers also benefit from the more structured training workflows. \textbf{SBERT-miniLM}, for instance, improves from 35.00\% to 73.3\% in Hit@1 and from 41.34\% to 76.9\% in MRR. Instruction-tuned models like \textbf{voyage-l2-instruct} show a similar trend, with Hit@1 improving from 37.86\% to 61.3\% and MRR from 42.44\% to 69.3\%. The performance of \textbf{multi-ada-002} also increases substantially across all metrics, particularly in Recall@50 (from 36.99\% to 62.2\%). These gains highlight that instruction-tuned and multivector models are particularly adept at capturing task relevance when the workflows follow consistent formats and terminology.

Interestingly, the improvement is most dramatic for models that previously underperformed. \textbf{ada-002}, which struggled on the general dataset (MRR = 3.94\%), performs considerably better on training workflows (MRR = 20.6\%). Similarly, \textbf{LLM2Vec} and \textbf{GritLM-7b} improve by 15–20 percentage points in most metrics. This suggests that the semantic clarity and reduced variability in training workflows allows models to align query intent with workflow descriptions more effectively. The strong performance of \textbf{GPT-4o} (MRR = 67.9\%) and \textbf{Mistral-7B} (Recall@50 = 93.2\%) further indicates that reranking models can exploit the structured nature of the training workflows to achieve highly accurate results.

Overall, model performance is consistently stronger on curated training workflows across all categories: lexical, dense, and reranking models. This suggests that model effectiveness is not only a function of architecture or tuning but also heavily influenced by the structure, clarity, and semantic regularity of the underlying workflow corpus. These findings underscore the importance of metadata quality and domain consistency when designing systems for task-aware workflow retrieval.

\begin{table*}[ht]
\centering
\caption{Qualitative examples of retrieval using Mistral-7B. Retrieved workflows generally align with query intent, with some variation in ranking precision.}
\label{tab:case-study}
\resizebox{0.8\textwidth}{!}{
    \begin{tabular}{p{3.5cm} p{4.5cm} p{3cm} p{3.5cm}}
    \hline
    \textbf{Query} & \textbf{Top Retrieved Workflows} & \textbf{Gold Workflow(s)} & \textbf{Observations} \\
    \hline
    RNA-Seq differential expression using DESeq2 & DESeq2 pipeline; STAR-DESeq2 integration; edgeR alternative & DESeq2-based pipeline & Accurate match in top-1 \\
    \hline
    Metagenomic profiling of environmental samples & Kraken2-Bracken; MetaPhlAn3; HUMAnN3 & MetaPhlAn3 workflow & All top results relevant \\
    \hline
    Find SNPs in whole-genome resequencing data & FreeBayes variant caller; GATK HaplotypeCaller; bcftools call & GATK-based variant calling & Partially aligned, GATK not top-1 \\
    \hline
    Single-cell RNA-seq clustering with Seurat & Scanpy-based pipeline; Seurat integration; Cell Ranger preprocessing & Seurat-based clustering & Relevant workflows present but not ranked first \\
    \hline
    Visualize differential abundance in microbiome & Volcano plot; LEfSe; QIIME2 taxonomic barplots & LEfSe pipeline & Retrieval partially aligned with intent \\
    \hline
    \end{tabular}
}
\end{table*}

\subsection{Semantic Improvement Through Rerankers}

To illustrate the impact of LLM-based reranking, we present a case study using the query posted on the Galaxy User Forum\footnote{\url{https://help.galaxyproject.org/}}. 

\textit{"How can I incorporate an annotated bacterial genome for RNA sequencing analysis using galaxy?"}\footnote{\url{https://help.galaxyproject.org/t/new-annotated-bacterial-genome/8894}}.

Table~\ref{tab:qual-case-study} compares results before and after applying Mistral-7b reranking. The original retrieval (via \texttt{voyage-l2-instruct}) surfaces workflows that are partially related to sequencing or annotation but lack alignment with the full task intent. For example, the top-ranked workflows include general quality control procedures, RAD-seq population genomics, or species identification workflows, none of which directly address bacterial genome annotation in the context of RNA-seq.

In contrast, Mistral-7b reranking significantly improves semantic alignment with the query, bringing to the top workflows that explicitly involve loading bacterial genomes, annotating them with tools like Prokka, integrating annotations with genome viewers (e.g., JBrowse), or assembling transcriptomes with StringTie. Several reranked results directly reference steps critical to the user’s goal, such as combining genome annotations with RNA-seq data and updating annotations with lncRNAs. While not all reranked workflows mention bacterial genomes explicitly, they more clearly support the integration of annotation and RNA-seq analysis workflows in Galaxy.

These results illustrate how generative LLM reranking helps bridge the semantic gap in dense retrieval by aligning workflow descriptions more closely with multi-step, domain-specific user intents.

To further explore the semantic alignment between user queries and retrieved workflows, we conducted a qualitative analysis across a diverse set of task-driven queries. Table~\ref{tab:case-study} summarizes representative examples using the Mistral-7B reranker. For each query, we compare the top retrieved workflows to a known gold-standard workflow and provide observations on ranking quality.

We observed that retrieval effectiveness varied based on the specificity and structure of the query. For well-scoped queries like \textit{"RNA-Seq differential expression using DESeq2"}, the top result precisely matched the intent, retrieving workflows built around DESeq2 or similar tools. Likewise, metagenomics queries returned consistently relevant workflows, often including multiple correct options within the top ranks.
However, in more complex or underspecified cases, such as \textit{"Single-cell RNA-seq clustering with Seurat"} or \textit{"Visualize differential abundance in microbiome"}, the retrieved results were relevant but inconsistently ranked. Some workflows included necessary steps or tools (e.g., Seurat, QIIME2, LEfSe) but were not prioritized first. We also noted partial mismatches in cases where closely related tools (e.g., FreeBayes instead of GATK) were retrieved for variant calling tasks.
These patterns suggest that while LLM rerankers like Mistral-7B improve task-aware alignment, further gains may be possible through enhanced query rewriting, tool disambiguation, or hybrid ranking strategies. Specifically, models may benefit from improved handling of tool-level synonyms, alternate pipelines, and multi-tool configurations.

\subsection{Retrieval Latency and Deployment Considerations}
We examined the average latency for each retrieval model across five representative task queries. As shown in Table~\ref{tab:latency-cost}, lexical models such as \textbf{TF-IDF} and \textbf{BM25} offer sub-millisecond response times, achieving an average latency of just 0.5 milliseconds per query. These models are ideal for real-time, interactive applications due to their negligible overhead and fast computation. \textbf{Fuzzy Matching}, though heuristic-based, is slightly slower (19.6 ms) but still practical for lightweight applications.

In contrast, \textbf{dense retrievers} present moderate latency increases while offering superior semantic capabilities. \textbf{SBERT-miniLM} and \textbf{BGE-base} operate under 5 milliseconds, making them attractive for low-latency semantic search. \textbf{DPR} remains usable at 5.6 ms but underperforms semantically. Commercial dense embedding APIs such as \textbf{ada-002} (1.33s) and \textbf{multi-ada-002} (0.79s) introduce measurable latency due to remote API calls, though they offer better retrieval performance. \textbf{VoyageAI's voyage-l2-instruct} strikes a reasonable balance with 0.32s latency and high top-k accuracy, positioning it as a viable real-time retriever.

LLM-based \textit{rerankers}, however, incur significantly higher latency. \textbf{GPT-4o} achieves strong relevance ranking at the cost of $\sim$1 second per rerank, while \textbf{Mistral-7B}, running locally, requires nearly 18.6 seconds due to large model size and autoregressive decoding. These models are therefore better suited for batch reranking or offline use cases. Table~\ref{tab:latency-cost} also contrasts API costs (where applicable), revealing a performance-cost-latency tradeoff. While \textbf{GPT-4o} and \textbf{voyage-l2-instruct} offer the highest retrieval performance, their monetary and temporal costs must be weighed against deployment requirements in real-world Galaxy interfaces. 

\begin{table}[ht]
\centering
\caption{Latency and API cost comparison of retrieval models.}
\label{tab:latency-cost}
    \resizebox{0.6\linewidth}{!}{
        \begin{tabular}{lccc}
        \hline
        \textbf{Model} & \textbf{Latency (s)} & \textbf{API Cost}\\
        \hline
        TF-IDF & 0.0005 & Free \\
        BM25 & 0.0005 & Free  \\
        Fuzzy & 0.0196 & Free \\
        SBERT-miniLM & 0.0022 & Free \\
        BGE-Base & 0.0040 & Free \\
        DPR & 0.0056 & Free \\
        ada-002 & 1.3257 & \$$\sim$0.0005/query \\
        voyage-l2-instruct & 0.32 & \$$\sim$0.00125/query \\
        multi-ada-002 & 0.79 & \$$\sim$0.00065/query \\
        GPT-4o Reranker & 1.0095 & \$$\sim$0.026/query  \\
        Mistral-7B Reranker & 18.59 & Free \\
        \hline
        \end{tabular}
    }
\end{table}

\begin{figure}
    \centering
    \includegraphics[width=0.8\linewidth]{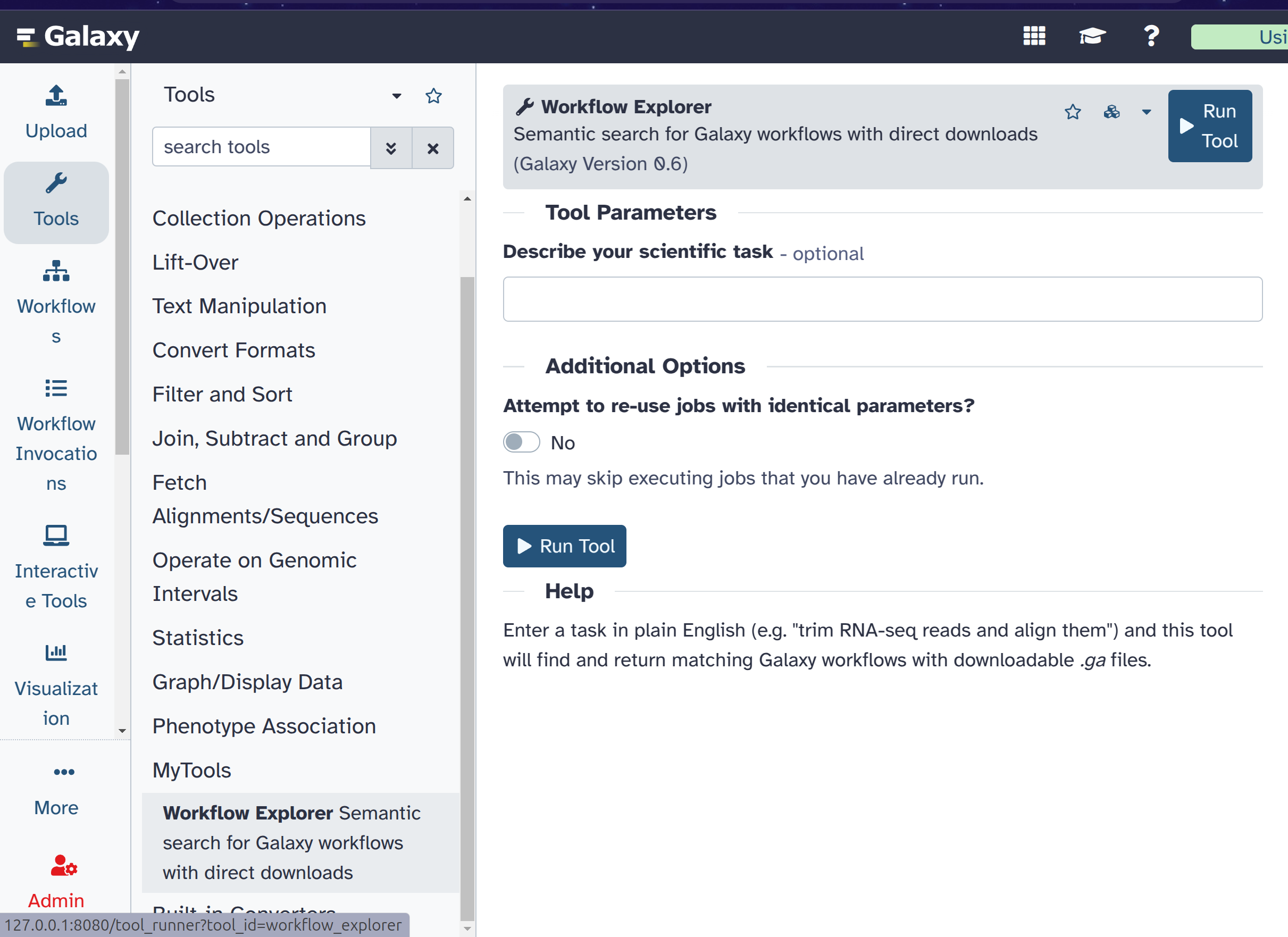}
    \caption{WorkflowExplorer tool interface in Galaxy: users provide a high-level natural language query.}
    \label{fig:interface1}
\end{figure}

\begin{figure}
    \centering
    \includegraphics[width=0.8\linewidth]{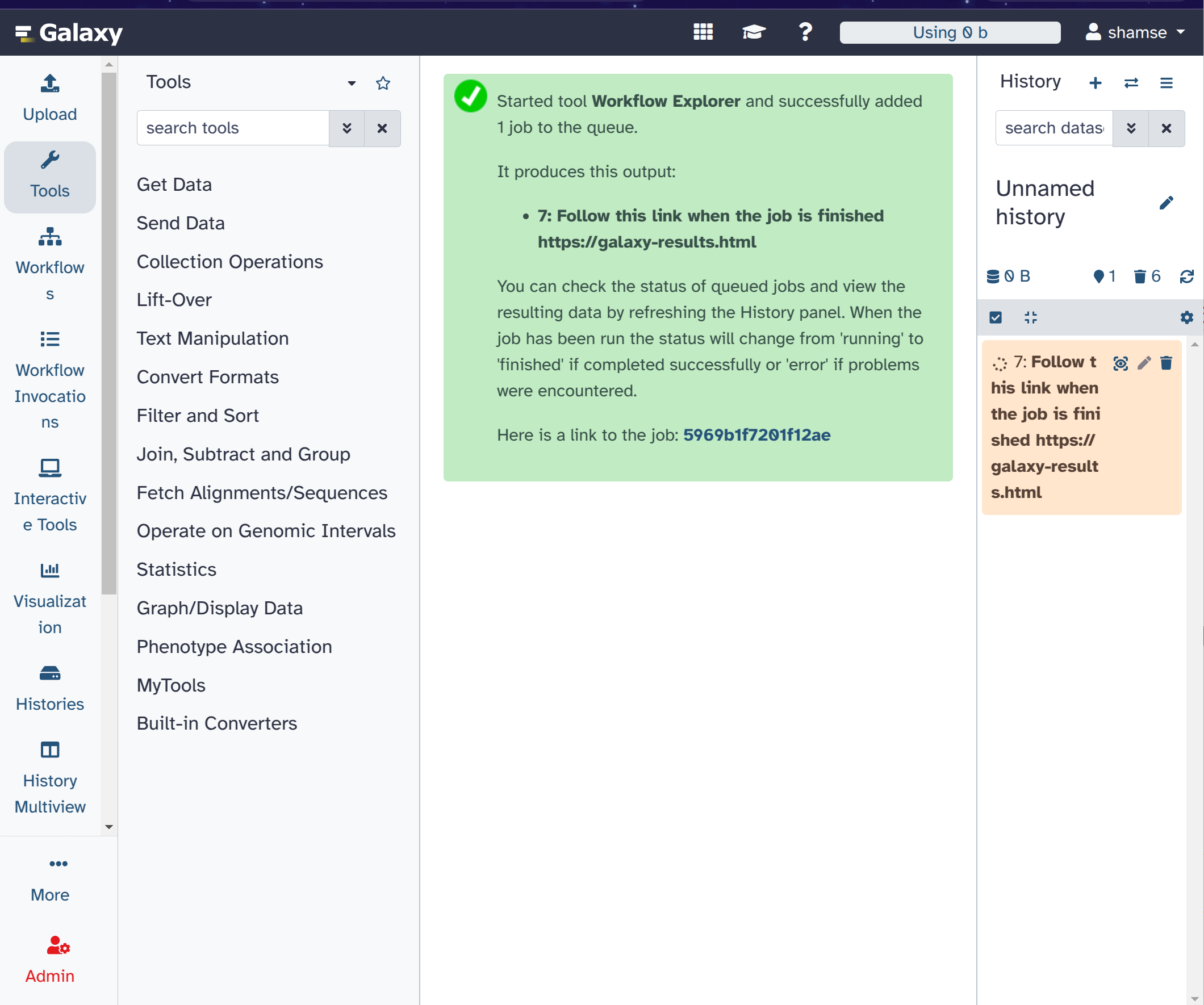}
    \caption{Galaxy job confirmation view with a hyperlink to the HTML results page.}
    \label{fig:interface2}
\end{figure}

To further validate the practicality of our proposed framework, we developed a prototype system and integrated it within a local instance of the Galaxy platform. This system serves as a proof-of-concept for semantic workflow retrieval powered by dense embeddings and LLM-based reranking, directly supporting user interaction through a task-oriented query interface.

As shown in Figure~\ref{fig:interface1}, users can access the tool from the Galaxy sidebar under \textit{MyTools}. The interface provides a minimal input form where users can describe their scientific task in natural language (e.g., ``\textit{How can I incorporate an annotated bacterial genome for RNA sequencing analysis using Galaxy?}''). Upon clicking ``Run Tool'', the input query is processed through a semantic retrieval pipeline that uses dense vector search with the \texttt{voyage-large-2-instruct} model, followed by reranking via a local \texttt{Mistral-7B-Instruct} model.

\begin{figure}
    \centering
    \includegraphics[width=0.8\linewidth]{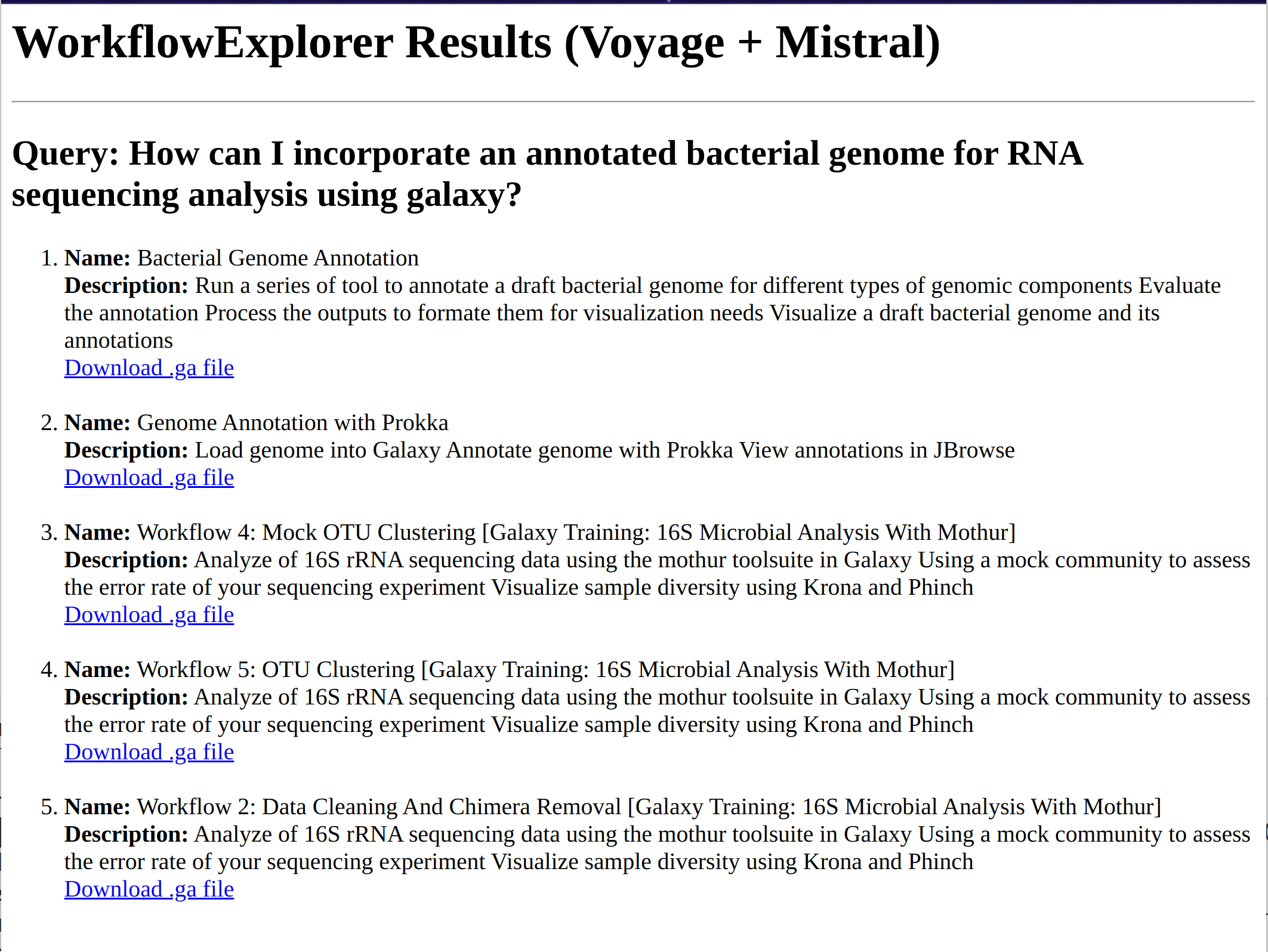}
    \caption{HTML output showing top-matched workflows, descriptions, and direct \texttt{.ga} file downloads.}
    \label{fig:interface3}
\end{figure}

Once the job is executed, Galaxy displays a history entry with a link to the result page (Figure~\ref{fig:interface2}). Clicking this link opens a formatted HTML output (Figure~\ref{fig:interface3}) showing the top-$k$ matched workflows. For each result, the tool name and description are presented, along with a direct download link to the associated \texttt{.ga} file. These files can be imported into Galaxy's workflow editor for further customization or reuse.

By embedding this functionality within Galaxy, the system allows users to locate semantically relevant workflows without requiring knowledge of tool names or syntax, thereby lowering the barrier to entry for new users and enhancing workflow discoverability and reuse.
\section{Prototype System Integration}

\section{Discussion}

\subsection{Retrieval Effectiveness and Model Behavior}

Our results reveal that dense vector models especially instruction-tuned models like \texttt{voyage-l2-instruct} and \texttt{multi-ada-002} consistently outperform lexical baselines in retrieving relevant Galaxy workflows. This supports the hypothesis that semantically rich, intent-aware embeddings are better suited for scientific queries \cite{glater2017intent}, which often involve descriptive, task-oriented language. We also observe that reranking with GPT-4o or Mistral-7B significantly improves top-rank accuracy and overall relevance, indicating that generative LLMs can capture better task relevance beyond what vector similarity can offer.

Interestingly, small dense retrievers such as \texttt{SBERT-miniLM} offer a competitive performance-cost tradeoff, achieving strong metrics while being lightweight and easy to deploy. In contrast, general-purpose models like \texttt{ada-002} perform poorly, underscoring the importance of domain or instruction alignment. This suggests that model architecture alone is not sufficient. Alignment with the target task and query domain are crucial for effective retrieval.

\subsection{Dataset Influence: General vs. Curated Workflows}

The difference in retrieval performance between general-purpose workflows and curated training workflows is striking. Across nearly all models, performance improved on the curated dataset, particularly in Hit@1 and MRR. This suggests that consistent structure, clear annotations, and reduced ambiguity in training workflows help models better align queries with relevant descriptions. These findings underscore the critical role of metadata quality and query clarity in the effectiveness of scientific workflow retrieval systems~\cite{song2024metadata}. They further suggest that platforms such as Galaxy could benefit from the integration of optional curation workflows or automated semantic tagging pipelines to improve the discoverability and usability of published workflows.

\subsection{Real-World Implications}

For researchers in bioinformatics, educators, and clinical scientists, the ability to retrieve analytical workflows based on high-level task descriptions is significantly more effective than relying on tool-specific wordings or remembering exact workflow titles. \textit{WorkflowExplorer} addresses this semantic gap by enabling task-aware retrieval, allowing users to locate relevant workflows even in the presence of vocabulary mismatch or domain-specific jargon. In educational contexts, this facilitates more effective learning by allowing students to explore workflows through conceptual queries, rather than relying on memorized syntax or toolchains. For the users, it streamlines reuse by helping domain experts identify pre-existing workflows that match multi-step experimental objectives, thereby accelerating data analysis and reducing redundant effort. Overall, \textit{WorkflowExplorer} lowers the cognitive and technical barriers associated with scientific workflow discovery, while promoting transparency and reusability.

\section{Threats to Validity}

\textbf{Internal Validity.} The quality of the benchmark dataset directly impacts model evaluation. Although we used a rigorous LLM-based pipeline to generate queries and constructed ground truth labels using topic-aware semantic filtering, the relevance assignment process may still introduce noise. While domain experts reviewed and filtered the queries for alignment and coverage, some mislabeling or overfitting to specific tool descriptions may persist.

\textbf{External Validity.} Our evaluation is confined to workflows within the Galaxy platform and may not generalize to other SWfMSs such as Taverna or Nextflow. Moreover, our benchmark includes workflows primarily in the domains of genomics and metagenomics. Retrieval performance might vary in other domains (e.g., proteomics or cheminformatics) where workflows are more structurally complex or sparsely annotated.

\textbf{Construct Validity.} The use of standard IR metrics (Hit@k, Recall@k, MRR) ensures comparability with prior work, but they may not fully reflect the real-world utility or satisfaction of retrieved workflows. For example, a top-ranked workflow might be technically relevant but practically unusable due to missing annotations or poor documentation. Incorporating user-centered evaluation, such as usability studies or expert satisfaction ratings, is left for future work.

\textbf{Implementation Validity.} The latency and effectiveness of our reranking step depend on the specific deployment conditions. While we offer both local (Mistral-7B) and API-based (GPT-4o) options, real-time integration into production-scale Galaxy instances may require further optimization, especially regarding inference costs, system load, and API quotas.

\textbf{Reproducibility.} We provide implementation details, source code, and datasets; however, the availability of commercial models (e.g., GPT-4o, voyage-l2-instruct) may restrict full replication. To mitigate this, our pipeline supports model substitution with open-access alternatives, and we release precomputed embeddings for public reuse.

\section{Conclusion and Future Work}

In this paper, we introduced \textsc{WorkflowExplorer}, a task-aware retrieval framework that LLMs for semantic search over Galaxy workflows. Our system integrates dense retrievers and rerankers (e.g., \texttt{GPT-4o}, \texttt{Mistral-7B}) to map user-specified natural language tasks to topically relevant workflows, thereby overcoming limitations of keyword-based retrieval in Galaxy. We constructed two benchmark datasets, one based on real Galaxy workflows and another from curated GTN tutorials—and conducted extensive experiments across 13 retrieval and reranking models. Results demonstrate that LLM-powered retrieval significantly improves Hit@k, Recall@50, and MRR compared to traditional sparse retrievers, albeit with higher latency and cost in some cases.
To facilitate real-world adoption, we also implemented a prototype Galaxy interface that allows users to query using natural language, preview ranked workflows, and download reusable \texttt{.ga} files directly. Our visual interface aims to reduce the cognitive burden on novice users and promote workflow reuse across scientific communities.

\textbf{Future Work.} In future, we aim to integrate user feedback signals (e.g., click-through, reuse) to fine-tune retrieval models via reinforcement learning. Second, we plan to extend our evaluation with user studies involving domain scientists to assess perceived relevance and usability. Third, to support even broader domains, we are exploring automatic annotation and summarization of unlabelled workflows using LLMs. Finally, we envision extending \textsc{WorkflowExplorer} into a multi-modal assistant that incorporates execution history, tool dependencies, and provenance information to provide context-aware workflow recommendations.

\bibliographystyle{IEEEtran}
\bibliography{ref}
\end{document}